\documentstyle[12pt]{article}
\newskip\humongous \humongous=0pt plus 1000pt minus 1000pt
\def\caja{\mathsurround=0pt} \def\eqalign#1{\,\vcenter{\openup1\jot
\caja	\ialign{\strut \hfil$\displaystyle{##}$&$
\displaystyle{{}##}$\hfil\crcr#1\crcr}}\,} \newif\ifdtup

\def\gtap{\raisebox{-.4ex}{\rlap{$\sim$}} \raisebox{.4ex}{$>$}}

\def\Nto1{\raisebox{-1ex}{\rlap{\tiny $\;\;N \to 1$}} \raisebox{0ex}
{$\;\;\;\,\to\;\;\;\,$}}

\def\frac#1#2{ {{#1} \over {#2} }}

\def\as{\alpha_S}
\def\sas{\sqrt{\alpha_S}}
\def\beq{\begin{equation}}
\def\eeq{\end{equation}}

\def\np#1#2#3{Nucl.\ Phys.\ B#1 (19#3) #2}
\def\pl#1#2#3{Phys.\ Lett.\ #1B (19#3) #2}
\def\pr#1#2#3{Phys.\ Rev.\ D #1 (19#3) #2}
\def\prep#1#2#3{Phys.\ Rep.\ #1 (19#3) #2}

\def\zp#1#2#3{Zeit.\ Phys.\ C#1 (19#3) #2}

\def\figcap{
        {\bf\large Figure Captions\markboth
        {FIGURECAPTIONS}{FIGURECAPTIONS}}\list
        {Fig. \arabic{enumi}:\hfill}{\vspace {4 mm}\settowidth
        \labelwidth{Figure 999:}
        \leftmargin\labelwidth
        \advance\leftmargin\labelsep\usecounter{enumi}}}
 \relax

\begin{document}
\par \vskip 10mm
\begin{center}
{
\Large \bf
QCD and high energy hadronic interaction:\\
Summary talk (theory)}\footnote{
Research supported in part by MURST, Italy.
Talk given at the XXVIIth Rencontre de Moriond,
March 1993 Les Arcs, France}
\end{center}
        \par \vskip 3mm \noindent
\begin{center}
        \par \vskip 3mm \noindent
        { GIUSEPPE MARCHESINI}\\
        \par \vskip 2mm \noindent
        Dipartimento di Fisica, Universit\`{a} di Parma,\\
        INFN, Gruppo Collegato di Parma, ITALY.\\
\end{center}
\par \vskip .7 true cm
\begin{center} {\large \bf Abstract} \end{center}
\begin{quote}
Results presented in perturbative QCD are reviewed.
The topics discussed include:
struction functions,
heavy flavour production,
direct photon production,
inclusive production at LHC/SSC,
small $x$ physics,
QCD jets and
intermittency.
For reference to the data in this talk see the summary presented
by J.E.  Augustin.
\end{quote}
\vspace*{\fill}
\renewcommand{\thefootnote}{\fnsymbol{footnote}}

\vskip .3 true cm
\noindent
{\bf 1. Struction functions.}
\vskip .2 true cm

New data
on deep inelastic structure functions by NMC and CCFR
experiments, together with a reanalysis of EMC and SLAC data
have been used by the Durham and CTEQ groups \cite{Durham}
to construct a new 1992 set of structure function
parameterization. This topic has been summarized by W.K. Tung.
The importance of the new data can be appreciated by observing that
in the region of $x$ between $10^{-1}$ and $10^{-2}$
the gluon density $G(x,Q)$ in the two sets are very similar.
However, if one \cite{BQ} takes into account also the new CDF data
on $b$-production a modification of $G(x,Q)$ in the
region of $x \sim 0.05$ is required.
The gluon density will soon be constrained by Hera measurements
down to values of $x\sim 10^{-4}$.
\vskip .3 true cm
\noindent
{\bf 2. Heavy flavour productions in next-to-leading order.}
\vskip .2 true cm

These are typical short distance processes due to the large mass of
the heavy quark $M_Q \gg \Lambda_{QCD}$. The canonical formula for
the heavy flavour production in $pp$ hadronic collider has the form
\beq \label{coll}
\sigma^{(pp)}(s, \cdots) \simeq
\int dx_1 f^{(p)}_{a_1}(x_1,\mu) \int dx_2 f^{(p)}_{a_2}(x_2,\mu)
\, \as^2(\mu)\, \hat \sigma_{a_1,a_2}(x_1x_2s,\mu,\cdots) \, ,
\eeq
where $\hat\sigma_{a_1,a_2}$ is the distribution for
the elementary process of heavy quark production from parton
$a_1$ and $a_2$ at the factorization scale $\mu$. This distribution
can be evaluated order by order in perturbative QCD
\beq
\sigma_{a_1,a_2} = \sigma^{(0)}_{a_1,a_2} +
 \as(\mu)\, \sigma^{(1)}_{a_1a_2} + \cdots \, ,
\eeq
where the Born $\hat\sigma^{(0)}$ and one loop contribution
\cite{ES,NDE} $\hat\sigma^{(1)}$ have been computed.
The functions $f^{(p)}_{a_i}(x_i,\mu)$ are the parton $a_i$ densities with
fraction $x_i$ of the incoming proton energy.
By using the one loop result for the elementary distribution
$\hat\sigma_{a_1a_2}$ and the one loop kernel for the parton
density evolution equation one computes the production cross
section to the so called next-to-leading order (NLO).

The physical cross section $\sigma^{(pp)}$ does not depend on the
factorization scale $\mu$ (here assumed equal to the renormalization
point). This means that the $\mu$ dependence in the parton densities
has to be compensated by the $\mu$ dependence in the elementary
distribution $\as^2(\mu)\hat \sigma_{a_1a_2}$.
Since these functions are known only up to first loop order, the
computed physical cross section has a residual $\mu$ dependence
which can be used to detect the importance of higher order
corrections and to estimate a ``theoretical error''.
Higher order corrections contain powers of $\ln \mu/Q$ with $Q$ the
hard scale of the process. This implies that, in order to minimize
higher order corrections one should anyway assume $\mu$ of order
$Q$. Typically one explores the $\mu$ dependence taking $\mu=a\,Q$
with $a=1/2-2$.
The strong $\mu$ dependence observed when taking only the leading term
(the Born contribution $\hat \sigma^{(0)}_{a_1a_2}$  and the first
order parton density evolution) is substantially reduced in the NLO
calculations.
However, even in the NLO calculations, one still expects a sizable
$\mu$ dependence in the region of low $p_t$ for the heavy quark jet.
This is due to the fact that the one loop distribution
$\hat \sigma^{(1)}_{a_1a_2}$  contains a contribution from a diagram
with a gluon exchange which is not present in the Born approximation.
This contribution grows at small $x$ and becomes more important
than the Born term. Thus in the region of small $x$, which corresponds
to the region of low $p_t$, there is no compensation of the residual
$\mu$ dependence. Here the theoretical uncertainty it then large and
estimated to be a factor two.
The recent data on the low $p_t$ distribution are larger than the
NLO results, but they are compatible within the experimental and
theoretical uncertainties (see talk by V. Papadimitriou for possible
sources of overestimation of b-production data at low $p_t$).

The status of NLO calculations of heavy flavour production have been
summarized by P. Nason. The processes investigated and the associated
elementary distributions are:
1) single inclusive hadro-production process,
$a_1a_2 \to Q +X$ with X any system of partons.
\cite{NDE};
2) single inclusive photo-production process,
$\gamma g \to Q +X$ and $\gamma q \to Q +X$.
\cite{NE};
3) tow-inclusive hadro-production,
process $a_1a_2 \to Q \bar Q +X$ \cite{MNR};
4) tow-inclusive photo-production,
$\gamma q \to Q \bar Q + X$ and $\gamma g \to Q \bar Q +X$.
\cite{FMNR};
5) single inclusive lepto-production,
$\gamma^* g \to Q +X$ and $\gamma^* q \to Q +X$.
\cite{LRSV};
6) resummation of higher order contributions at small $x$
\cite{CCH,HFEP}.

P. Nason presented the calculation of tow-inclusive photo-production.
In this process there are two contributions:
the ''photon point-like'' and ''photon resolved'' term.
In the first one the photon is directly involved in the elementary
process ($\gamma g \to Q \bar Q +X$ and $\gamma q \to Q \bar Q +X$),
while in the second term the parton distribution inside the photon
is involved.
This process, presently studied  at Hera, gives informations
both on the gluon density at $x$ as small as $10^{-4}$
and on the parton densities into the photon.

NLO calculations of inclusive lepto-production has been discussed
by E. Leanen. In this case the ``photon resolved'' contribution is
negligible for large off-shell photon mass $Q^2 >4 $ GeV$^2$.
Therefore this process is suitable for obtaining constraints on the
gluon distribution at small $x$. The calculation can be done to NLO by
using the one loop elementary distribution and the NLO parton
densities.
For instance the $F_2$ or $F_L$ structure functions for heavy flavour
leptoproduction can be cast in the form
\beq
F(x,Q^2,M_{Q\bar Q})=
\int dx f_a(z,\mu)C_a\left( \frac x z,M_{Q\bar Q},Q^2,\mu \right)
\eeq
where $C_a$ are the Wilson coefficient function computed from the
elementary distribution of the processes
$\gamma^* g \to Q +X$ and $\gamma^* q \to Q +X$.

The structure function $F$ should be independent of the
factorization/renormalization scale $\mu$. The importance of higher
order contributions can be studied by analyzing the $\mu$
dependence of the NLO results. For values of $x$ away from $x=0$ and
$x=1$ the result is quite independent of $\mu$  for $\mu$ near the
hard scale $\mu=a \sqrt{4M_Q^2+Q^2}$  with $a=1/2-2$.

For large $x$ there are large higher order corrections involving
powers of $\ln(1-x)$ which are due to soft gluon emission.
The leading contributions $\as^n \ln^n(1-x)$ (and some next to leading
corrections) can be easily resummed by taking into account the
coherence of the soft gluon radiation \cite{IR}.

For small $x$ there are higher order corrections involving powers of
$\ln x$. As discussed in Ref.~\cite{CCH}, the leading contributions $\as^n
\ln^n x $ can be resummed by using the ``Lipatov'' anomalous dimension
(see later) and the generalized $k_t$-factorization theorem
which is based on the fact that at small $x$ one should take into
account the gluon and the photon off-shell mass in the elementary process
$\gamma^* g \to Q +X$.
Therefore the heavy flavour structure function for small $x$ is
given by  an expression of the type
\beq
F(x,Q^2,M_{Q\bar Q}) \simeq
\int dx d k^2_t \, {\cal G}(z,k^2_t)
\, \hat \sigma\left(\frac x z,M_{Q\bar Q},Q^2, k^2_t\right)
\eeq
where $ \hat \sigma$ is the off-shell Born distribution with
$-Q^2$ and $-k^2_t$ the photon
and gluon virtual mass squared (only physical polarizations contribute
at small $x$ thus this quantity is gauge invariant).
In this formula the function ${\cal G}(x,k^2_t)$
is the generalized gluon density giving the probability
(per unit of $\ln x$) of finding
a gluon at longitudinal momentum fraction $x$ and transverse
momentum $k_t$.
Integrating this distribution over $k_t<\mu$ one obtains
the gluon density $G(x,\mu^2)$
\beq\label{int}
G(x,\mu)=\int^{\mu^2} dk^2_t {\cal G}(x,k^2_t)
\eeq
By studying  $\hat \sigma(z,M_{Q\bar Q},Q^2, k^2_t)$ at Hera energy
one has \cite{HFEP} that the hard scale is typically
$\mu^2=4M_Q^2+Q^2$.
However at small $Q^2$ or at c.m. energy $W \sim M_Q$,
the dynamical suppression in $k_t$ is at a significantly
smaller scale.

In the conventional calculation, the heavy flavour cross section is
obtained by convoluting the on-shell elementary cross section and the
gluon density $G(x,\mu^2)$. This procedure has two main effects:
(i) for $k_t^2<\mu^2$ the elementary cross section is overestimated by
its on-shell value;
(ii) the `tail' of the cross section at $k_t^2>\mu^2$ is ignored.
Asymptotically, the second effect dominates and the
cross section is expected \cite{CCH} to be larger than
the conventional on-shell Born approximation.
At subasymptotic energies the first effect is important
and one overestimates the cross section.
Monte Carlo simulation based on the coherent branching
algorithm \cite{HFEP}
predicts that the b-quark leptoproduction cross section at Hera
energy is lower than the one obtained by a conventional one-loop
calculation \cite{EK}.

\vskip .3 true cm
\noindent
{\bf 3. Direct photon production}.
\vskip .2 true cm

The importance of this process is based on the fact that ``photons do
not hadronize''. Thus hadronization, a corrections beyond perturbative
QCD, does not affect the calculation. In principle this is then a nice
way to measure, for instance, the gluon density.
However there are various points which should be understood before
a complete reliable calculation could be used.
J. Qiu presented the status of the calculation to NLO. This includes
NLO hard elementary distribution, NLO parton densities and NLO photon
fragmentation function.
This last quantity is the first source of uncertainty.
In purely inclusive data the photon fragmentation function gives
the most important contribution \cite{Fon}.
{}From perturbative QCD we can compute the dependence of
the photon fragmentation function on the factorization scale $\mu$,
which has to be taken of the order of hard jet transverse energy.
This is given by the Altarelli-Parisi evolution equation and the
kernel is know to NLO.
However, as in the case of parton densities, to compute the photon
fragmentation function we need to implement the initial condition, i.e.
a distribution at a fixed scale $\mu_0$ which should be obtained
from the data (see Ref.~\cite{ZT} for a discussion at LEP).

For experimental reasons the measure is done by selecting ``isolated
photons'', i.e. by requiring that within a given angle $\delta$
the photon is accompanied by an energy fraction less than
$\epsilon$. The isolation has the effect of reducing by a large
fraction the size of the contribution from the photon fragmentation
function, thus reducing the size of the uncertainty.
However, we know that in perturbative QCD when we select events
as with these isolation requirements, the distribution contains double
logarithmic contributions $\as \ln \delta \ln \epsilon\,$ giving
rise to Sudakov form factors. These quantities are well understood
in perturbative QCD.
Since these Sudakov corrections are large, one needs to define an
algorithm to isolate the photon which must be the same in experiments
and in the NLO perturbative QCD resummation.
At present the CDF data agree with the QCD calculation within the
errors and the presence of the residual theoretical uncertainties.

\vskip .3 true cm
\noindent
{\bf 4. Inclusive production at LHC/SSC}
\vskip .2 true cm

With the advent of very energetic hadron colliders a precise
estimation of inclusive production of neutral particles is important
in order to pin down signals due to Higgs particles or new physics.
For large transverse momentum $p_t$ the distribution can be
computed in perturbative QCD.
A consistent NLO calculation of inclusive $\pi^0$ production has been
presented by P. Chiappetta.
The distribution can be written as a convolution
\beq
\sigma^{(pp)}(x,...)
=
\int \frac {dz}{z}
\hat \sigma_{pp \to a }
\left(\frac x z ,\as(\mu),\mu, ...\right)
D^{\pi^0}_a(z,\mu)
\eeq
where $ \hat \sigma_{pp \to a }$ is the inclusive distribution
for emitting a parton $a$ from the two incoming protons.
The function $D^{\pi^0}_a(z,\mu)$ is the $\pi^0$ fragmentation function.
The determination of $ \hat \sigma_{pp \to a }$ can be done to NLO
accuracy by using the one loop elementary distributions
\cite{ES,ACG}
and the NLO parton densities.
As described in the previous case of prompt photon production,
at present, the NLO evolution kernel for the fragmentation function
is known.
However we need to implement $D^{\pi^0}_a(z,\mu)$ at some scale $\mu_0$.
In the calculation presented the distribution has been evaluated
by using the Monte Carlo simulation HERWIG at $\mu_0=30$GeV and
compared with Cello data.
The distribution is then evolved with NLO corrections at ISR and UA2
energies and checked. Finally a prediction is given at LHC.
The theoretical uncertainty can be analyzed by changing the
factorization scale $\mu$ of the order of the hard scale given by
the large $p_t$ of the emitted $\pi^0$.
The theoretical uncertainty is estimated to be within a factor
two.

Nice data for total photoproduction cross section at Hera are expected.
This is a typical ``soft physics'' problem since no hard scale is involved.
First attempts of computing the part of the cross section due to
hard collisions with emission of high $p_t$ jets have been presented
by I. Sarcevic and G. Schuler. The uncertainty here is mostly on the
extrapolation of perturbative QCD results to low $p_t$ scale.

\vskip .3 true cm
\noindent
{\bf 5. Small $x$ physics}
\vskip .2 true cm

One of the most arduous problems in perturbative QDC is the
the analysis of processes involving incoming hadrons in the region
$ \Lambda \ll Q \ll \sqrt s$
where $Q$ is the hard scale and $\sqrt s$ the c.m. energy
of the process. In this region the Bjorken variable
$x \simeq Q^2/s$ is small.
An example of these processes is heavy flavour leptoproduction at Hera,
where the heavy quark mass $M$ sets the scale of the hard process.
This region, which is now available at Hera for the first time,
has been the subject of intensive studies in perturbative QCD.
Recent new results allow one to formulate the problem in
a complete form at least to leading order. Namely one is now able
to attempt the analysis not only of the behaviour of the cross
sections, of its size, but also of the structure of the radiation
associated to these processes.
The main results have been reported by L. Lipatov, M. Ryskin.

1.  {\it Lipatov equation}.

In perturbative QCD the field of small $x$ physics started \cite{FKL}
with the fundamental works by Fadin, Kuraev and Lipatov (FKL)
in the framework of soft type of physics, namely, without any hard scale.
Originally they considered a theory with a massive gauge field and
they evaluated the total cross section as function of $s$ by resumming
terms of type $\as^n \ln^n s/m_g^2$ with $m_g$ the gluon mass.
One finds that the various terms can be arranged within a ladder type of
kinematical topology of the multi-gluon emission.
The contribution of the exchange of a gluon with a squared sub-energy
$\hat s$ grows proportionally to $\hat s$. It was found that
resumming the mentioned contributions one should ''reggeize'' the
gluon by substituting
\beq
\label{FKL}
\eqalign{
&
\frac {\hat s}{m_g^2} \equiv
\frac 1 z \to
\frac 1 z \Delta_{FKL}(z,k_t,m_g) \,,
\cr&
\Delta_{FKL}(z,k_t,m_g) =
exp \left\{
-\bar \as \int_z^1 \frac {dz'}{z'}
\int_{m_g^2}^{k_t^2}
\frac {d q'^2_t} {q'^2_t}
\right\}
=
\left( \frac {\hat s}{m_g^2}\right)^{-\bar \as\ln k_t^2/m_g^2}\,
}
\eeq
where $\bar\as= C_A\as/\pi$ and $k_t$ is the transverse momentum
of the exchanged gluon.
The total cross section is obtained by integrating over $k_t$
a distribution ${\cal G}(x=m_g^2/s,k_t)$ which satisfies the Lipatov
equation. The presence of the ''reggeized'' gluon turned out to
be crucial in order to satisfy a fundamental property of cancellation
of singularities for $k_t \to 0$, so that one can avoid the
cutoff given by the gluon mass $m_g$.

{\it 2.  All loops anomalous dimension}

A further fundamental contribution to this field
is due to L.V. Gribov, Levin and Ryskin in Ref.~\cite{GLR}.
Here the problem was formulated in the framework of hard processes
and the Lipatov equation was rederived and discussed.
The distribution ${\cal G}(x,k_t)$, corresponds to a
generalized gluon density giving the probability (per unit of
$\ln x$) of finding a gluon at longitudinal momentum fraction $x$
and transverse momentum $k_t$.
Integrating this distribution over $k_t<\mu$ one finds the gluon
density at scale $\mu$ as in (\ref{int}) (see also Ref.~\cite{IR}).
The gluon density is given in term of the space-like
anomalous dimension  $\gamma^S_N(\as)$
(the limit $x \to 0$ corresponds to $N \to 1$, where $N$ is the
energy moment index).
The leading contributions in $\gamma^S_N(\as)$ are given by an
expansion in powers of $\as/(N-1)$.
The first terms of the ``Lipatov'' anomalous dimension are
\begin{equation}
\label{gammae}
\gamma^S_N (\as) =
\frac {\bar \as} {N-1}
+ 2 \zeta_3 \left(\frac {\bar \as} {N-1}\right)^4
+ 2 \zeta_5 \left(\frac {\bar \as} {N-1}\right)^6
+12 \zeta_3^2 (\frac {\bar \as} {N-1})^7
+ \cdots
\end{equation}
where $\bar \as = C_A \as/\pi$ and
$\zeta_i$ is the Riemann zeta function.
There ere are no leading terms of order $\as^2$, $\as^3$, and $\as^5$.
Although each term is singular only at $N=1$, this expansion
develops a square root singularity at $N = 1+ (4 \ln 2 ) \bar \as$.
The presence of this singularity at $N>1$ implies
that the behaviour of the gluon density for $x \to 0$ is more
singular than that given by any finite number of loops.
For fixed $\as$ and small $x$ the behaviour of the one loop
gluon density is
\begin{equation}
\label{sf1}
G^{(1)}(x,Q) \sim \exp \sqrt{a\ln(1/x)}\;,
\;\;\;\;\;\;\;\;\;\;\; a = 4\bar\as\ln(Q^2/Q^2_s)\; .
\end{equation}
By summing the all loop result in Eq.~(\ref{gammae}) one finds
instead the following behaviour
\begin{equation}
\label{sfall}
G^{(all)}(x,Q) \sim x^{-p}\;,\;\;\;\;\;\;
p=(4 \ln 2)\bar\as \simeq 0.5 \; ,
\end{equation}
which is much more singular for $x\to 0$.

{\it 3. Unified branching}

Recently \cite{CCFM} the technique of soft gluon factorization
theorems \cite{IR} has been extended to the field of small
$x$ physics.
It was found that the resummation of the leading $\as^n \ln^n x $
terms gives rise to multi-gluon emission with a ladder type of kinematical
diagrams. One finds coherence phenomena arising from interference
among emitted gluons. The result is that the emission can be described
by a branching process in which gluons are successively emitted in a
phase space ordered in angles. This is the same phase space valid also
in the complementary region in which $x \to 1$. Therefore one can
formulate a branching process which is valid in all region of $x$ even
for $x \to 0$ or $1$.
One finds then a unified picture for parton emission which for finite
$x$ gives rise to the usual light-cone evolution equation while
for small $x$ one finds the Lipatov equation. This unified branching
process can be used to construct a Monte Carlo simulation program
\cite{HFEP} in which many features can be studied quantitatively.

In spite of the quite different behaviours of the gluon densities
(\ref{sf1}) and (\ref{sfall}), it turns out that the all-loop and
the conventional one-loop formulations give similar results.
This is partially due to the fact that the first correction to the
one-loop expression of the anomalous dimension is to order $\as^4$.
Thus the steeper behaviour of the gluon density function is seen only
for very low $x$.
Moreover, as pointed out in the talk of M. Ryskin, the steeper
behaviour for $\as$ running is even more asymptotic than for fixed $\as$.
This is due to the presence of the cutoff in the exchanged transverse
momenta $k_{t}>Q_0$. Although this condition is asymptotically negligible,
it has some effect in reducing the evolution of the
branching in the first steps. As a result the distribution at small
$x$ is somewhat reduced.

In the study of heavy flavour leptoproduction \cite{HFEP},
the most important
differences between the improved all-loop branching and the
conventional one-loop branching are seen in the final state
gluon distributions.
These differences arise from the
additional phase space available for primary gluon emission in the
all-loop evolution. The number of emitted gluons is enhanced,
especially at small $x$ and large angles, i.e. in the low-rapidity region.
At present, these differences are small compared with uncertainties due
to our lack of knowledge of the input gluon distribution. This underlines
the importance of determining the gluon density experimentally
down to the lowest possible values of $x$.

It may be surprising that one can extend the soft gluon theorems from
the region $x \to 1$ to the region $x \to 0$.
In the first case one has that all emitted gluons are soft, while in
the second case one has that the exchanged gluons are soft instead.
Actually for small $x$ one finds that $x$ plays the role of an
infrared cutoff. Due to the LNK theorem all gluons
emitted with energy fractions smaller than $x$ do not give singular
contributions and therefore can be neglected to leading order.
Thus the limit $x\to 0$ is again a soft physics problem.
The most important new effect in the region of small $x$ is the
presence of new virtual singularities negligeble for finite $x$.
They can be exponentiated and give rise to a new form factor
called the ''non-Sudakov'' form factor. It is given by
\beq
\Delta_{ns}(z,q_t,k_t) =
exp \left\{
-\as\int_z^1\frac {dz'}{z'}\int_{(z'q_t)^2}^{k_t^2}
\frac {d q'^2_t} {q'^2_t}
\right\}
\eeq
where $z$, $k_t$ are the energy fraction and transverse momentum of
the exchanged gluon, while $q_t$  is the transverse momentum of the
emitted gluon.
Comparing with (\ref{FKL}) we see that this form factor corresponds to
the ``reggeization'' of the gluon in the Lipatov calculation.

The calculation done so far are for the leading terms
$\as^n \ln^n x$ or in terms of moments $\as^n/(N-1)^n$.
It has been pointed out in this meeting that the next to leading
calculations of terms $\as^n/(N-1)^{n-1}$ are of crucial importance.
First of all they are required in order to control the argument of
the running coupling. Observe that changing the scale one finds
\beq
\frac {\as(\rho Q)} {N-1}= \frac {\as(Q)} {N-1}
-\beta_0 \ln \rho \frac {\as^2(Q)}{N-1} +\cdots
\eeq
where $\beta_0$ is the first coefficient of the beta function.
The present level of accuracy does not allow one to control these
corrections.
Even most importantly one finds that in the limit $x\to 0$ the
generalized gluon density ${\cal G}(x,k_t)$
depends only on $k_t$ while the hard
scale itself is lost. This is just the Regge type of factorization.
In this way however one does not have the property of renormalization
group, a property typical of hard scattering physics. This can be
obtained only by an accurate control of the hard scale $Q$. The
scale enters in the calculation for $x \to 0$ only to next to
leading order. It was then pointed out in this meeting the urgency
of these NLO  calculations.

{\it 4. Unitarity corrections}

The growth of the gluon density at small $x$ in (\ref{sf1}) and
(\ref{sfall})
are violating unitarity. There are then two important questions:
(i) how unitarity is restored;
(ii) at what value of small $x$  should one find experimental
indications of this violation.
To obtain such indications one needs a model of unitarity
restoration. The most practical model has been suggested by Gribov
Levin and Ryskin in their important Ref.~\cite{GLR}. They
proposed the GLR equation for the gluon density at small $x$
\beq
\frac{\partial^2G(x,Q)}{\partial \ln x \partial \ln Q^2}
= \frac {\as C_A}{\pi} G(x,Q)
-\frac {81 \left(\as G(x,Q)\right)^2} {16R^2Q^2}\,,
\eeq
where unitarity is implemented by the second term in the r.h.s.
The equation with only the first term is the Altarelli
Parisi equation for small $x$ (the anomalous dimension is given by
the first term in (\ref{gammae})). The second term, coming from
higher twist Feynman diagram contributions, slows the growth
of the parton distribution and leads to the saturation of the parton
density. The observability of the saturation has been discussed by
Ryskin and depends on the value for the parameter $R$. A uniform
distribution of partons corresponds to $R\sim 1$fm.
If the partons are clustered in hot spots a value of $R\sim 0.2$fm
would be more appropriate.
An unambiguous observation of unitarity saturation seems to be
very difficult.
As discussed before, even at the very small values for $x$ which
will be explored at Hera it will be difficult to detect the difference
between the one and all loops behaviour in (\ref{sf1}) and
(\ref{sfall}).

As discussed by Ryskin, recently \cite{BLR} it has been found that
there are Feynman diagrams contributions at the required level of
accuracy which are not included in the GLR equation.
The modifications required are under study.
These contributions seems to be small since they are non planar
thus suppressed by $1/C_A^2$.

A more systematic way to implement unitarity has been discussed by
L. Lipatov. Taking into account that in the small $x$ region the
multi-gluon emission takes place as a two dimensional random walk in
the transverse momentum,
Kirschner, Lipatov and Szymanowski propose to describe the
multi-gluon emission by an effective field theory in two dimension.
This theory should resum all relevant higher twit diagrams,
includes absorption and should satisfy unitarity.
It is interesting that one can study the solution of this effective
lagrangian with the modern techniques of conformal field theory.

One of the necessary consequences of this study is the presence of
the odderon. This has been discussed by B. Nicolescu
which reported a recent estimation of the odderon singularity.
Recalling that the anomalous dimension in (\ref{gammae}) is at
$N=N_P=1+(4\ln2)\as C_A/\pi\simeq 1.5 $, one finds that the odderon
singularity is at $N \gtap 1+0.13\,(4\ln2)\as C_A/\pi\, \gtap \, 1.07 $.
This would predict that the difference
$\sigma_{pp}-\sigma_{p \bar p}$ grows with energy.

C.I. Tan reported a nice attempt \cite{LT} to merge the hard
perturbative QCD results at small $x$ and some important
phenomenological properties of soft physics.
The model is based on the following two features:
(i) in hard processes the multi-gluon emission at small $x$ takes place
as a diffusion process in $t=\ln(k_t/\Lambda)$;
(ii) in soft physics the emission takes place as a diffusion process
in impact parameter $\bf b$.
In the model then one assumes (i) diffusion in $t$ for any $\bf b$;
(ii) diffusion in $\bf b$ only in the soft physics region $t<t_0$
with $t_0$ a cutoff corresponding to a transverse momentum of the
order of a GeV.

\vfill
\newpage

\vskip .3 true cm
\noindent
{\bf 6. QCD jets}.
\vskip .2 true cm

Hard processes are characterized by the presence of jets. They
are not elementary degree in the lagrangian, so they need an
operatorial definition, a ''jet-finding'' algorithm.
The main ingredients for a ''jet-finding'' algorithm are the rule
for clustering particles and the jet resolution.
Within perturbative QCD one should make on the algorithm
the following requirements:
(i) it should involve the proper QCD hard scale which is dictated by
coherence. In $e^+e^-$ the scale is the total c.m. energy $Q$,
the photon virtuality $Q$ in deep inelastic scattering,
$E_t$ in hadron-hadron processes with large transverse
energy processes;
(ii) it should be easy to use theoretically (and experimentally
of course). In particular it should allow higher order calculations
and resummation when required such as in the case of small jet
resolutions;
(iii) it should be not very sensitive to hadronization corrections.
Recall that the fact that colour is neutralized locally \cite{AV}
in the transverse momentum of partons is consistent
with perturbative QCD.
Phenomenologically the size of the
hadronization could be estimated by Monte Carlo simulations involving
colour coherence and local colour neutralization.

The ''Snowmass--accord'' algorithm, introduced few years ago for
defining jets in hadron colliders, has been summarized by S. Ellis.
This has been extensively used for instance by the CDF collaboration.
For each particle or calorimeter cell
with rapidity, azimuthal angle and transverse energy
$\eta_i$, $\phi_i$ and $E_{ti}$, one defines a  radius
\beq r_i  =\sqrt{(\eta_i-\eta_J)^2+(\phi_i-\phi_J)^2}
\eeq
where $\eta_J$, $\phi_J$  and $E_{tJ}$ are the corresponding variables
for a jet.
The particle or calorimeter cell belong to the jet if $r_i$ is smaller
than a given resolution $R$ which is typically of order one.

The importance of this algorithm is that it allowed for the first
time the possibility of a comparison of data with NLO calculations
of $p_t$ distributions \cite{ES,ACG}.
In the presentation made by S. Ellis some difficulties where
pointed out:
a separation cut $E_0$ is needed and this makes
difficult to study the problem of merging two jets and changing
the resolution;
when the resolution $R$ becomes small the distributions develop
$\ln R$ powers which become large and they need to be resummed;
the algorithms is not suited for higher order resummations;
underlying radiation, which can not be studied by perturbative QCD,
is not easy to isolate by this algorithm.

S. Catani and M. Seymour discussed recent theoretical developments
of a ''jet-finding'' algorithms applied to Lep, Hera, and hadronic
colliders hard processes.
This is similar to the ''Jade'' algorithm which was originally
proposed for $e^+e^-$ annihilation. Schematically it can be defined
as follows:
For each pair of particle momenta $p_i$ and $p_j$ one constructs a
''resolution'' variable $y_{ij}$ (to be defined later).
Consider the two momenta with minimum value of $y_{ij}$.
If $y_{ij} < y_c $ the two particle momenta are combined to give a
single momentum. The parameter $y_c$ is the jet resolution parameter.
The clustering is stopped when non pair of momenta satisfies the
bound. The resulting number of jet of a given event is then a function
of $y_c$.
The key quantity in this algorithm is the proper definition  of the
''resolution'' variable $y_{ij}$.
One needs a definition which would allow simple
higher order calculations and $\as^n\ln^n y_c$ resummations when the
resolution becomes small.
To this end one should takes into account that perturbative QCD
is characterized by the presence of collinear singularities
for small angle $\theta_{ij}$ between two partons and
infrared singularities when a parton become soft.
This observation is the basis of the recently proposed
the $k_t$/Durham algorithm and resolution variable $y_{ij}^{k_t}$.
To see the main difference with the old ''Jade'' resolution variable
consider the case of two partons with ordered energy $E_i \ll E_j$
and small relative angles.
In this limit, the Jade and $k_t$ resolution variables tends to
\beq
\eqalign{
&
y_{ij}^{Jade} =\frac{2E_iE_j(1-\cos\theta_{ij}) } {Q^2} \Rightarrow
\frac {E_iE_j}{Q^2}\theta^2_{ij}
\cr&
y_{ij}^{k_t} =\frac{2Min(E^2_i,E^2_j)(1-\cos\theta_{ij}) } {Q^2}
\Rightarrow \left( \frac {E_i\theta_{ij} }{Q}\right)^2
\simeq \left(\frac{k_{ti}^{(j)}}{Q}\right)^2
}
\eeq
where $k_{ti}^{(j)}$ is the transverse momentum of the soft parton $i$
with respect to parton $j$.
We see then that in the limit of soft and collinear partons the  $k_t$
algorithm exposes the proper kinematical variable angle time
soft energy.
For small resolutions, this fact allows one to exponentiate
and resum the leading and next to leading powers of $\ln
y_c$. Moreover the $k_t$ algorithm, as expected, is less
sensitive to hadronization corrections, as can be phenomenologically
estimated by Monte Carlo simulations.
At small resolution parameter the exponentiation is obtained from
soft gluon factorization theorems which are at the basis of the coherent
branching process.

M. Seymour discussed the extension of the $k_t$-jet algorithm to hard
processes with incoming hadrons, i.e. at Hera or hadron colliders.
Also in these processes a prominent feature is the presence of hard
QCD jets.
Together with these jets one has also radiation coming from the
remnants of the incoming hadron.
This radiation is not described by perturbative QCD since no hard
scale is here involved.
In order to measure and compute distributions of QCD jets involving
the hard scale one has to find an operative definition of the jets
which does not involve soft physics radiation.
To this end one observes that the characteristic phenomenological
feature of soft physics is that the radiation is bounded to be within a
fixed values of transverse energy independent of the hard scale.
Consider for instance the jets emitted in deep inelastic process
at Hera with $Q$ the photon virtuality.
The $k_t$ algorithm in this process is preceeded by an
algorithm which is intended to isolate the soft physics radiation.
This isolation algorithm is defined at a scale $E_c$ with
$\Lambda \ll E_c \ll Q$.
{}From the emitted particles or calorimeter cells with energies $E_i$
with angles $\theta_{ip} $ with respect to the incoming proton and
with relative angle $\theta_{ij}$ between two particles
one computes the resolution parameters
\beq
Y_{ip}= \frac {2E_i^2}{E^2_c}(1-\cos\theta_{\theta_{ip}})\,,
\;\;\;\;\;
Y_{ij}= \frac {2Min(E_i^2,E_j^2) }{E^2_c}(1-\cos\theta_{\theta_{ij}})
\,.
\eeq
We have one of the following situations:
(i) the minimum resolution parameter is $Y_{ip}<1$.
In this case particle $i$ is associated to the remanent;
(ii) the minimum resolution parameter is $Y_{ij}<1$.
In this case the two particles are combined into a single
''particle'';
(iii) no resolution parameter is smaller than one.
In this case the combination ends.
After this algorithm is applied one has two sets of particles,
for a given cutoff $E_c$: particles which belong to the
proton remanent and hard jet particles we can analyse by the
same $k_t$-algorithm introduced in $e^+e^-$ annihilation.
The algorithm here depends on two resolution
parameters $E_c$ and $y_c$.
The same algorithm can be generalized to the case of two incoming
hadrons.

On the theoretical point of view, the important feature of this
algorithm is that no subtraction of underlying event is necessary thus
avoiding one of the most important difficulties of the
Snowmass--accord. The resummation at small resolution of the large
$\ln y_c$ powers can be done as in $e^+e^-$ to NLO accuracy.
Moreover one can show that collinear singularities are factorized.
This is important in order to take into account the NLO corrections
in the elementary distributions and parton densities.
In the analysis of jets great importance plays the scale of the
hard process, as emphasized by S. Catani.
In general one finds that the proper scale for jet emission is
given by coherence.

\vskip .3 true cm
\noindent
{\bf 7. Intermittency}
\vskip .2 true cm

A nice summary of intermittency has been presented by A. Bialas.
The most characteristic feature of intermittency is that the
multiplicity moments have fractal dimensions.
Consider two cones of angular apertures $\delta$ and $\Delta>\delta$.
The moments of particles emitted within a cone have fractal dimension if
\beq
\label{delt}
F_2(\delta) \equiv \frac{<n(n-1)>_{\delta}}{<n>^2_{\delta}}
= \left( \frac{\Delta}{\delta} \right)^{f_2}
F_2(\Delta) \,.
\eeq
This behaviour can be generalized to all multiplicity moment ratios
$F_s(\delta)$.
The importance of this quantity in describing dynamical properties
is clear if we observe that for $\delta \to 0$ Eq.~(\ref{delt}) implies
the singular behaviour $F_2(\delta) \sim \delta^{-f_2}$.
Similarly also the two particle correlation is singular when the
particles become parallel.
We have then that the fractal dimension $f_2$ is a fundamental
parameter for describing in a unified way various dynamical features.
This become even more interesting in the hypothesis proposed by
Fialkowski that $f_2$ is universal, independent of
particles and reactions.

The prediction of fractal dimension within perturbative QCD has been
described in the contributions of W. Ochs, J. Meunier and I. Dremin.
One finds in general
\beq\label{nu}
F_s(\delta)/F_s(\Delta) \sim
\left( \frac {<n>_\delta}{<n>_\Delta}\right)^{\nu(s)}
=\left( \frac{\delta}{\Delta} \right)^{\omega(s)}
\eeq
where $\nu(s)$ and $\omega(s)$ are anomalous dimensions \cite{DMO}
related to the time--like gluon anomalous dimension for $N \to 1$ which
to leading order is
\begin{equation}
\label{tgamma}
\gamma^T_N(\as) = \sqrt{ \bar \as +
\left(\frac{N-1}{4}\right)^2 }\,
-\frac{N-1}{4}
\end{equation}
where $\bar \as = {C_A \as}/{2 \pi}$.
For running $\as$ one find \cite{DMO}
\beq
\nu(s) \simeq \frac{s-\omega(s)}{1-\sqrt \tau}\,,
\;\;\;\;\;\;
\omega(s) \simeq s\sqrt \tau\,(1-\frac {\ln \tau}{2s^2})\,,
\;\;\;\;\;
\tau=\frac{\ln (E\delta/\Lambda)}{\ln (E\Delta/\Lambda)}
\eeq
The coherent branching algorithm allows one to compute all these
quantities to leading order.
We must however be aware of the following limitations on the
accuracy of the calculation:

\noindent
(i) in perturbative calculation we require a large scale which in this
case give a limitation on the smallest angle to consider
$ E\delta \gg \Lambda$,
with $E$ the jet energies. Small angle could then be reached at large
jet energy;

\noindent
(ii) all multiplicity moments obtained by the coherent branching
algorithm are to leading and next to leading order \cite{FW}.
Since soft gluon emission is
very singular, it turns out that multiplicity moments have the
expansion parameter $\sqrt \as$ rather then $\as$.
Usually only the first corrections in $\sas$ are known.
There is one case in which we know higher order terms in the $\sas$
expansion. This is the KNO scaling function
\beq
\eqalign{
&
<n(n-1)\cdots (n-s+1)>= g_s(\sas) <n>^s\,,
\cr&
g_s(\sas)=g_s(0)(1-s\sas+ \cdots c_m(s\sas)^m \cdots)\,,
}
\eeq
where all coefficients of $s\sas$ powers have been evaluated
\cite{DD} and one finds
large modification to  leading order KNO formula;

\noindent
(iii) hadronization corrections should be estimated.
They should factorize in the ratios $F_s(\delta)$
and should be small for $E\delta$ much larger than
some hadronization scale of the order of a GeV.
It is interesting to note that, due to the universality of the
fractal dimension one could phenomenologically study the behaviour of
$F_s(\delta)$ even at smaller $E\delta$. This could be used in
order to learn some constraints on the hadronization process

\par \vskip 0.5 true cm

\end{document}